# Improving the Imaging Performance of Microwave Imaging Systems by Exploiting Virtual Antennas

Xinhui Zhang, Naike Du, Jing Wang, Andrea Massa, *Fellow, IEEE*, and Xiuzhu Ye, *Senior Member, IEEE*

*Abstract*—Starting from the observation that the correlation coefficient defined by the scattered field data tested by two adjacent antennas decreases with the noise, it turns out that the imaging performance can be improved by adding non-redundant scattered field information through more measuring antennas. However, adding more measuring antennas faces practical challenges such as the limited antenna space, high experimental expenses, and a prolonged data collection time. Therefore, the frequency-domain zero-padding (FDZP) interpolation method is proposed to acquire scattered field data on more virtual antennas. To process the data, a linear inversion algorithm based on the modified Born approximation (MBA) and the nonlinear subspace-based optimization method (SOM) are used to image scatterers of moderate and high contrasts, respectively. The effectiveness and the reliability of the proposed approach are then assessed against synthetic data, semi-experimental data from a full-wave simulation software, and experimental data.

## I. INTRODUCTION

In recent years, microwave imaging has gained significant attention due to its non-ionizing radiation, super-resolution imaging, and quantitative reconstruction capabilities [1]-[6]. This technique aims to determine the shape, the position, and the distribution of the constitutive parameters of unknown scatterers within the domain of interest (DOI) by processing the scattered field collected outside the DOI [7]. Its applications span various fields such as biomedical imaging [3][8], microscopic imaging [9][10], geological exploration [11][12], and non-destructive testing and evaluations (NDT/NDE) [13][14]. To yield a microwave image, an inverse scattering problem has to be solved. This is a challenging task, owing to the inherent nonlinearity and ill-posedness. The solution methods for inverse scattering problems can be roughly categorized into two main classes: deterministic methods and stochastic techniques. Firstly, a cost function is commonly defined as the mismatch between measured and calculated data. Then the image of the DOI is obtained by iteratively or non-iteratively minimizing such a cost function. Deterministic methods comprise linear (e.g., the Born approximation algorithm (BA) [15] and the modified Born approximation (MBA) [16]) and nonlinear (e.g., the Distorted-Born Iterative Method (DBIM) [17], Contrast Source Inversion (CSI) method [18], and Subspace-based Optimization Method (SOM) [19]) approaches for weak and strong scatterers, respectively. Deterministic methods are characterized by a high convergence efficiency, but also they present the risk of being trapped into local minima or erroneous solutions, especially for the case of strong scatterers.

Unlike deterministic methods, stochastic techniques (e.g., Genetic Algorithms (GAs) [20], Particle Swarm Optimization (PSO) [21], and Differential Evolution (DE) [22]) prevent the solution from being trapped in local minima by generating a set of trial solutions with stochastic operators. However, they suffer from a high computational burden. To enhance the efficiency of stochastic methods, an innovative learned global optimization technique, which is based on an artificial intelligence-driven integration of evolutionary algorithms, has been recently proposed [23]. Otherwise, targeting to practical engineering applications, learning-by-examples (LBEs) techniques have been developed to address complex real-world problems in real-time scenarios [24][25]. More in detail, a computationally-efficient and accurate surrogate model of the implicit inverse operator is built by learning the relationship between the known input (i.e., the scattered field samples) and output (i.e., the contrast function) data during a preliminary off-line phase [24].

As for the data and to enhance the imaging performance, it is crucial collecting as much as possible non-redundant scattered field information [26], and the minimum number of non-redundant measurements can be determined by the degrees of freedom (DOFs) of the scattered field [26]. Generally, the scattered field is measured with a finite set of antennas (Fig. 1). In a noisy environment (Sect. III), increasing the number of antennas (NOAs) can improve the

Manuscript received xxxx, xxxx; accepted xxxx, xxxx. Date of publication , xxxx; date of current version xxxx, xxxx. This work was supported by the National Natural Science Foundation of China under Grant 61971036, the Fundamental Research Funds for the Central Universities and Beijing Nova Program.(*Corresponding Author: Xiuzhu Ye*).

X. Zhang, N. Du, J. Wang and X. Ye are with the ELEDIA Research Center (ELEDIA@BIT – BIT), School of Information and Electronics, Beijing Institute of Technology, Beijing 100081, China (e-mail: xiuzhuye@outlook.com).

A. Massa is with the ELEDIA Research Center (ELEDIA@UESTC - UESTC), School of Electronic Science and Engineering, University of Electronic Science and Technology of China, Chengdu 611731, China, also with the ELEDIA Research Center (ELEDIA@UniTN - University of Trento), DICAM- Department of Civil, Environmental, and Mechanical Engineering, 38123, Trento, Italy, also with the ELEDIA Research Center (ELEDIA@TSINGHUA-Tsinghua University), Haidian, Beijing 100084, China, and also with the School of Electrical Engineering, Tel Aviv University, Tel Aviv 69978, Israel (e-mail: andrea.massa@ing.unitn.it).

imaging performance. However, the use of a large NOAs in the imaging system would be hindered by practical constraints such the limited available installation space, high experimental expenses, and a prolonged data collection time [27]. Consequently, how to achieve better imaging performance with a reduced NOAs in a noisy environment holds paramount importance in the development of a practical imaging system. It is worth mentioning that increasing the NOAs is essentially an increase of the measured environmental information, that is not only on the scatterers under test, but also on the noise level and the "characteristics" of the measurement setup.

Given the limitations and the drawbacks of physically increasing the NOAs, researchers have investigated the virtual arrays based on an interpolation approach to reduce the NOAs, while simultaneously improve the imaging performance. This technique has been extensively used in medical imaging [27], ultra-wideband through-wall imaging [28], and synthetic aperture radar (SAR) imaging [29]. Recently, there has been a growing interest in virtual experiments to improve the contrast range of non-iterative inversion methods by the linear combination of the measured scattered field [30]-[32]. In certain cases, the required NOAs can be reduced by situating a conducting cylinder near the scatterer [33] or adjusting the distribution of the measurement antennas within a limited aperture [16]. Recently, the concept of "antenna pattern diversity" has been profitably introduced in indoor RF imaging. Since each node gathers multiple independent measurements, the number of measurement nodes is decreased [34]. Furthermore, machine learning has been exploited to solve the inverse scattering imaging under limited measurement aperture [35][36].

This study aims to investigate the impact of the NOAs on the performance of microwave imaging systems in noisy environments. Subsequently, an interpolation method based on the Frequency-Domain Zero-Padding (FDZP) is introduced to enhance imaging accuracy, while reducing the actual NOA thanks to the inclusion of virtual antennas. The linear Modified Born Approximation (MBA) method or the nonlinear Subspace-based Optimization Method (SOM) are applied to image scatterers with moderate or high contrasts, respectively. Finally, the effectiveness and the reliability of the proposed imaging approach are validated against synthetic and experimental data.

The main contributions of this paper lie in the following items:
(1) A proof that increasing the NOAs in noisy environments can effectively improve the imaging performance;
(2) The introduction of an interpolation method, based on FDZP, to profitably add virtual antennas that not only improves the imaging accuracy, but it also reduces the number of physical antennas;
(3) The definition of a criterion for setting the optimal and minimum NOAs in practical imaging systems;
(4) The assessment of the proposed approach based on virtual antennas against synthetic, semi-experimental, and experimental data.

The outline of the paper is as follows. In Section II, the theory of the forward and inverse scattering problems is detailed and the inversion methods used in the imaging process are briefly summarized. In Sect.III, the impact of the NOAs on the imaging performance is studied and an interpolation method is proposed to increase the number of scattered data, without adding more physical antennas. Reconstructions from synthetic (Sect. IV) and real-world experiments (Sect. V) are presented. Finally, some conclusions are drawn (Sect.VI).

## II. Formulation Of Forward and Inverse Scattering Problems

### A. Forward Scattering Problem

This paper focuses on 2-D microwave imaging under transverse magnetic (TM) wave illumination (Fig. 2). The unknown scatterer lies within the DOI, $D$, which is located in a background medium with permittivity $\varepsilon_b$ and permeability $\mu_b$. In practice, the transmitting antennas and the receiving antennas are usually composed of the same antenna array, so the transmitting/receiving antennas are evenly located in a circle outside $D$. There are $N_i$ transmitting antennas that illuminate the scatterer. For each incidence, the scattered field data are measured by $N_r$ receiving antennas and they are stored in the matrix $\bar{E}^{sca}$ of size $N_i \times N_r$. Each antenna can work in both transmitting and receiving mode that the number of transmitting and receiving antennas is equal ($N_r = N_i =$ NOAs). To numerically compute the scattered field, the DOI is discretized into $M$ uniform cells centered at ($\mathbf{r}_m, m = 1, 2, ..., M$). According to the Lippmann-Schwinger equation [7], the total electric field $\bar{E}^{tot}(\mathbf{r})$ in DOI can be expressed as

$$\bar{E}^{tot}(\mathbf{r}) = \bar{E}^{inc}(\mathbf{r}) + i\omega\mu_b \int_D g(\mathbf{r}, \mathbf{r}') \cdot \{-i\omega\varepsilon_b[\varepsilon_r(\mathbf{r}')-1]\bar{E}^{tot}(\mathbf{r}')\}d\mathbf{r}' \text{ for } \mathbf{r} \in D \quad (1)$$

where $\bar{E}^{inc}(\mathbf{r})$ is the incident electric field in the DOI, $g(\mathbf{r}, \mathbf{r}')$ is the Green's function, while $\varepsilon_r$ and $\omega$ are the relative permittivity in the DOI and the angular frequency, respectively. By using pulse basis functions and the point matching technique, the discretized Lippmann-Schwinger equation assumes the following matrix form:

$$\bar{E}^{tot} = \bar{E}^{inc} + \bar{\bar{G}}_D \cdot \bar{J} \quad (2)$$

where $\bar{\bar{G}}_D$ is the matrix modeling the electromagnetic interactions within the DOI, $\bar{J}$ is the vector of the induced current given by $\bar{J} = \bar{\bar{\xi}} \cdot \bar{E}^{tot}$, $\bar{\bar{\xi}}(m,m) = \xi(\mathbf{r}_m), m = 1, 2, ..., M$ being the contrast in the $m$-th cell

$$\xi(\mathbf{r}_m) = -i\omega\varepsilon_b[\varepsilon_r(\mathbf{r}_m)-1] \quad (3)$$

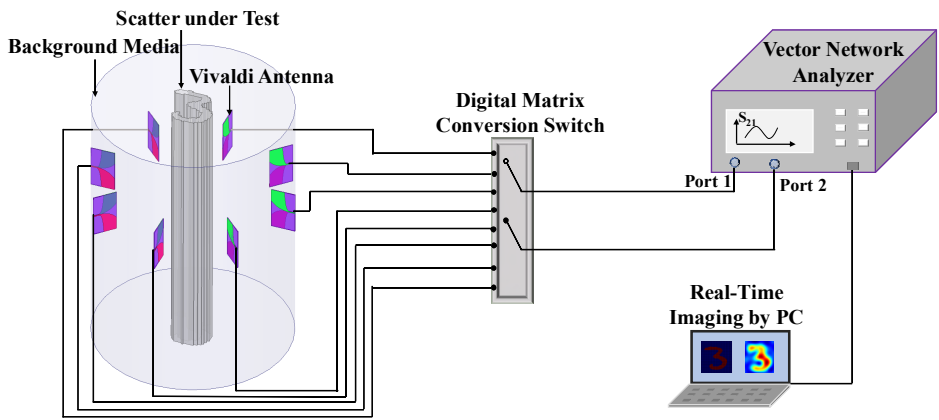

Fig. 1. Schematic diagram of a 2D microwave imaging system.

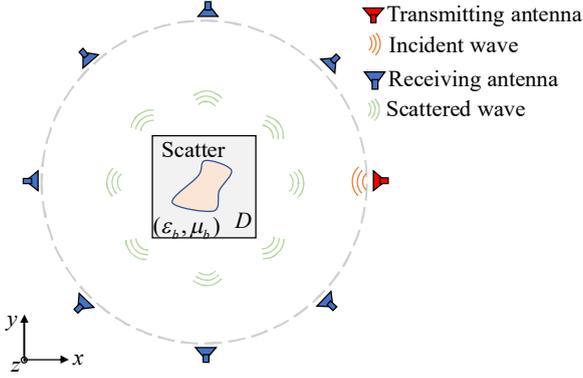

Fig. 2. Schematic diagram of 2D microwave imaging.

The scattered field at the $N_r$ receiving antennas is given by

$$\bar{E}^{sca} = \bar{\bar{G}}_S \cdot \bar{J} \quad (4)$$

where $\bar{\bar{G}}_S$ models the electromagnetic interactions between the induced current within the DOI and the receiving antennas.

To generate the synthetic scattered field data, the CG-FFT-MOM method is used [37].

*B. The FDZP Method*

To increase the information content from the measurements without adding more receiving/transmitting antennas (i.e., the NOAs), the FDZP method is adopted to "add" virtual antennas collecting non-redundant scattering data. The FDZP method [38] is based on the theory of discrete digital signal processing and it assumes that the "zero padding" in one domain (i.e., the discrete Fourier transform (DFT) domain of the scattered field) results in an increased sampling rate in the other domain (i.e., the scattered field). By using such a technique, the original NOAs is updated after interpolation to N (N > NOAs).

In order to detail the FDZP method, let $\bar{E}^{sca}_{NOA \times 1}$ be the original scattered field data vector. The following steps are then performed:

1) Perform DFT on $\bar{E}^{sca}_{NOA \times 1}$, namely $\tilde{E}^{sca}_{NOA \times 1} = DFT(\bar{E}^{sca}_{NOA \times 1})$;
2) If NOA is even (odd) then insert the zero vector $\mathbf{0}_{(N-NOA) \times 1}$ between the $\frac{NOA}{2}$ ($\frac{NOA+1}{2}$) and the $\frac{NOA}{2}+1$ ($\frac{NOA+3}{2}$) elements of $\tilde{E}^{sca}_{NOA \times 1}$ to yield the interpolated vector $\tilde{E}^{sca}_{N \times 1}$;
3) Perform the inverse DFT (IDFT) of $\tilde{E}^{sca}_{N \times 1}$ and scale it by a factor $\frac{N}{NOA}$ to obtain the interpolated scattered field data, namely $\bar{E}^{sca}_{N \times 1} = \frac{N}{NOA} IDFT(\tilde{E}^{sca}_{N \times 1})$.

Similarly, one can keep the number of receiving antennas unchanged, while increasing the number of transmitting antennas virtually. Finally, the result is that the scattered field data are stored in the vector $\bar{E}^{sca}_{N \times N}$.

*C. Inverse Scattering Methods*

In this paper, the reconstruction from the scattered data are performed with state-of-the art techniques, namely the linear MBA method or the nonlinear SOM when considering moderate or high contrast scatterers, respectively. For the sake of completeness, they will be summarized in the following.

Let us start by remembering that according to [7], the external matrix $\bar{\bar{G}}_S$ can be decomposed as $\bar{\bar{G}}_S = \bar{\bar{U}} \cdot \bar{\bar{\Sigma}} \cdot \bar{\bar{V}}$ by means of the Singular Value Decomposition (SVD) and the induced current can be expressed as $\bar{J} = \bar{\bar{V}} \cdot \bar{\alpha}$, where $\bar{\alpha}$ is a column vector containing the induced current coefficients. By substituting $\bar{J} = \bar{\bar{V}} \cdot \bar{\alpha}$ into (4), the $i$-th ($i = 1,...,M$) entry of $\bar{\alpha}$ turns out to be

$$\alpha_i = \frac{\bar{u}_i^H \cdot \bar{E}^{sca}}{\sigma_i} \quad (5)$$

where $\bar{u}_i$ and $\sigma_i$ are the $i$-th ($i = 1,\ldots,M$) left singular vector and the $i$-th ($i = 1,\ldots,M$) singular value of $\bar{\bar{G}}_S$, respectively. In practice, since measured scattered field data are usually corrupted by noise, it is convenient to avoid higher indexes that correspond to very small singular values, $\sigma_i$, otherwise the coefficients $\alpha_i$ would be wrongly predicted. Accordingly, only the first $L$ large singular values are used to compute the corresponding current coefficients $\alpha_i$.

On the other hand, due to the independence of the column vectors of $\bar{\bar{V}}$, the induced current $\bar{J}$ can be divided into two orthogonal and complementary subspaces, namely $\bar{J} = \bar{J}^+ + \bar{J}^-$, which are called the deterministic, $\bar{J}^+$, and the ambiguous, $\bar{J}^-$, parts, respectively. More specifically, the deterministic component is given by

$$\bar{J}^+ = \bar{\bar{V}}^+ \cdot \bar{\alpha}^+ \quad (6)$$

where $\bar{\bar{V}}^+$ is the matrix of the first $L$ right singular vectors of $\bar{\bar{G}}_S$ and the entries of the vector $\bar{\alpha}^+$ are the first $L$ coefficients of $\bar{\alpha}$. Similarly, the ambiguous part of the induced current is defined as $\bar{J}^- = \bar{\bar{V}}^- \cdot \bar{\alpha}^-$, where $\bar{\bar{V}}^-$ is the matrix of the remaining $M-L$ right singular vectors of $\bar{\bar{G}}_S$.

As for the MBA method, the scattered field within DOI generated by $\bar{J}^+$ is considered to approximate the total electric field in $D$

$$\bar{E}^{tot} = \bar{E}^{inc} + \bar{\bar{G}}_D \cdot \bar{J}^+ \quad (7)$$

Since the inverse scattering problem at hand is still highly ill-posed, the Tikhonov regularization technique is used to yield an MBA-image of the DOI. More in detail, the contrast $\bar{\xi}$ is reconstructed by minimizing the following cost function

$$f(\bar{\xi}) = \left\| \bar{\bar{G}}_S \cdot \bar{\bar{E}}^{tot} \cdot \bar{\xi} - \bar{E}^{sca} \right\|^2 + \gamma \left\| \bar{\xi} \right\|^2 \quad (8)$$

where $\|\cdot\|$ is the Euclidean norm and $\gamma$ is the regularization parameter selected according to the $L$-curve method [39].

Unlike the Born approximation method, which directly employs the incident field to approximate the total electric field within the DOI while disregarding multiple scattering effects within the scatterer, the MBA method accounts for the scattered field within the DOI generated by the deterministic component of the induced current, thereby partially considering the multiple scattering effects. This feature enables the MBA method to effectively image scatterers with moderate contrasts. While the MBA method uses only the deterministic part of the induced current, the exploitation of the ambiguous component of the induced current leads to the SOM. In this latter method, the cost function assumes the following expression

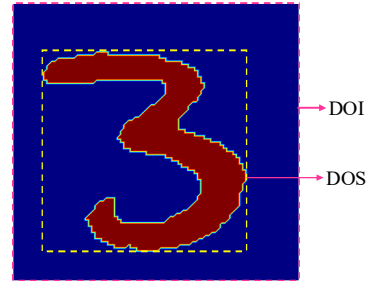

Fig. 3. Digital "3" model is embedded in DOI.

$$f(\bar{\xi}, \bar{\alpha}^-) = \frac{\left\| \bar{\bar{G}}_S \cdot \bar{\bar{V}}^- \cdot \bar{\alpha}^- + \bar{\bar{G}}_S \cdot \bar{J}^+ - \bar{E}^{sca} \right\|^2}{\left\| \bar{E}^{sca} \right\|^2} + \frac{\left\| \bar{\bar{A}} \cdot \bar{\alpha}^- - \bar{B} \right\|^2}{\left\| \bar{J}^+ \right\|^2} \quad (9)$$

where $\bar{\bar{A}} = \bar{\bar{V}}^- - \bar{\bar{\xi}} \cdot (\bar{\bar{G}}_D \cdot \bar{\bar{V}}^-)$ and $\bar{B} = \bar{\bar{\xi}} \cdot (\bar{E}^{inc} + \bar{\bar{G}}_D \cdot \bar{J}^+) - \bar{J}^+$. The minimization of (10) by using the conjugate gradient-type optimization scheme to determine the contrast $\bar{\bar{\xi}}$ and the ambiguous coefficients, $\bar{\alpha}^-$, of the induced current. Thanks to the integration of both the deterministic and the ambiguous components of the current, the SOM can image scatterers with high contrast, but at the cost of an increase of the computational time for the scattered data inversion.

III. NUMERICAL RESULTS BASED ON SYNTHETIC DATA (MOM)

In this section, the results from the processing of the synthetic scattered data generated with the MOM are presented to show the impact of the NOAs on the imaging performance. Such data are related to a two-dimensional inverse scattering system for biomedical imaging. More specifically, handwritten digital models have been assumed as scatterers. For example, Fig.3 shows the digital "3" model, which is embedded in a background medium with $\varepsilon_b = 37.725$ and loss tangent equal to 0.148 in a DOI of size 0.1 [m] × 0.1 [m], the DOS being the domain of the scatterer. The DOI has been illuminated by transmitting sources radiating ideal cylindrical waves at the frequency of 800 MHz. The transmitting/receiving antennas have been evenly distributed on a circle of 0.12 [m] radius centered in origin of the imaging system. As for the forward solver, the DOI has been uniformly partitioned into 50×50 square sub-domains, while the CG-FFT-MOM has been chosen to predict the scattered field samples in the measurement points. To prevent the inverse crime, the data inversion has been dealt with the same DOI, but it has been discretized into a 40×40 uniform square grid. Furthermore, to avoid bias due to the inversion method and considering the range of working of each reconstruction technique, the relative permittivity of the scatterers to be imaged with the MBA approach or the SOM has been randomly selected in the range [40, 50] or [50, 60], respectively.

The quality of the image reconstruction has been quantitatively assessed by using the structural similarity (SSIM) [40] index

$$SSIM(p_1, p_2) = [l(p_1, p_2)]^\tau [c(p_1, p_2)]^\upsilon [s(p_1, p_2)]^\varsigma \quad (10)$$

and the relative error (RE)

$$RE = \frac{\left\|\bar{\bar{\xi}}_o - \bar{\bar{\xi}}_r\right\|}{\left\|\bar{\bar{\xi}}_o\right\|} \quad (11)$$

where $p_1$ and $p_2$ represent true and reconstructed images, respectively, $l$, $c$ and $s$ are luminance function, contrast function, and structural function used to measure two images, respectively, and $\tau$, $\upsilon$ and $\varsigma$ are the parameters that adjust the relative weight of the three functions, and here they are all taken as 1. $\bar{\bar{\xi}}_o$ and $\bar{\bar{\xi}}_r$ are the true and reconstructed contrast, respectively. It is worth distinguishing that the contrast function $c$ measures the color contrast difference between two images, while the contrast $\bar{\bar{\xi}}$ corresponds to the relative permittivity within the DOI.

While both the SSIM and the RE fall by definition within the range between 0 and 1, higher/lower values of the first/second one indicates a better reconstruction.

*A. Estimation of the Minimum NOAs (DOF Theory)*

According to the DOFs theory, it is possible to estimate the minimum NOAs required for the imaging process. The number of the DOFs of a 2D problem is given by [26]

$$DOF = 2\beta a \quad (12)$$

where $\beta$ is the wave number in the background medium and $a$ is the minimum radius of the circle including the scatterer. Obviously, the minimum NOAs turns out to be equal to the DOFs when setting $a$ to the value of the radius of the smallest enclosure of the DOS.

Dealing with 1000 handwritten characters to model the scatterers and owing their variable size, the DOFs for these shapes are reported in Fig. 4. It turned out that the average number of DOFs across all these models is 9, so that the minimum NOAs for the imaging system has been set to this value. It is worth noting that also some real-world scenarios (e.g., human thoracic or brain stroke imaging) and not-only for synthetic experiments, there is a lot of *prior* knowledge on the size of the scatterer that makes it possible to estimate the minimum NOAs.

*B. Dependence of Imaging Performance on NOAs*

To assess the dependence of the imaging performance on the NOAs, the fields scattered by each of the 1000 models have been computed in the NOAs locations ranging from 9 to 20. These data have been blurred with a Gaussian white noise whose level is given by

$$nl = \frac{\left\|\bar{E}^{noi}\right\|_F}{\left\|\bar{E}^{sca}\right\|_F} \times 100\% \quad (13)$$

where $\left\|\cdot\right\|_F$ and $\bar{E}^{noi}$ are the Frobenius norm and the Gaussian white noise, respectively. Since the MBA method is more computationally efficient than the SOM, the former method has been used here by randomly setting the actual relative permittivity of the targets within $[40, 50]$.

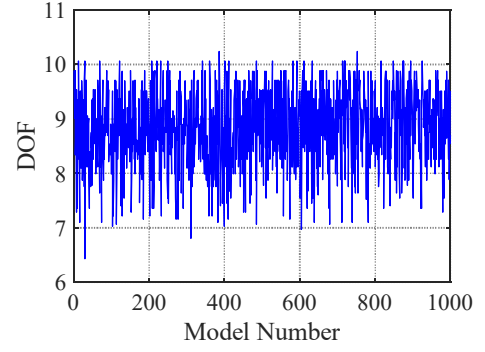

Fig. 4. Statistical DOF results of the 1000 handwritten digital models.

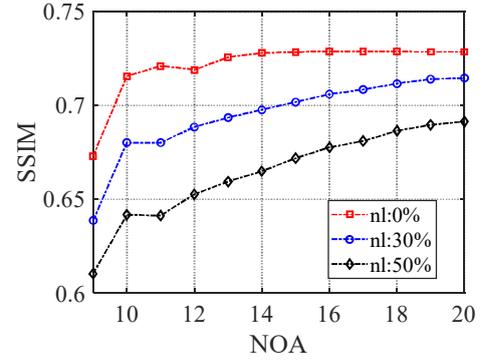

(a)

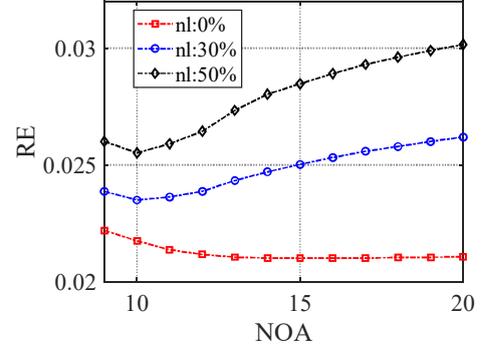

(b)

Fig. 5. Average SSIM and RE of imaging varying with actual NOA under different noise levels (a) SSIM; (b) RE.

Fig. 5 summarizes the results of the analysis by showing the behavior of the average value of the SSIM [Fig. 5(*a*)] and the RE [Fig. 5(*b*)] yielded by the MBA-inversions when varying the actual NOAs for different levels of noise. As it can be observed, increasing the NOAs improves the SSIM value for a fixed noise level. For instance, the SSIM grows from 0.673 up to 0.729 moving from NOA = 9 to NOA = 20 in the noiseless case. This also holds true when *nl* = 50% (SSIM = 0.610-NOA =9, SSIM =0.691-NOA =20). On the contrary, the RE cannot be decreased by increasing the NOAs and it slightly gets worse except for the noiseless case. However, the slight deterioration of the RE can be disregarded in comparison with the non-negligible improvement gained in the SSIM.

The reason why increasing the NOAs can improve the imaging accuracy, especially in noisy cases, has been then explored by analyzing the correlation coefficients [41]. By choosing the NOAs equal to 9 and 20, it has been assumed that the receiving antennas are sequentially disposed on the circumference from the first to the last *NOA*-th in a counter-clockwise direction. Fig. 6 plots the normalized correlation coefficient (NCC) of the scattered field data measured on two adjacent antennas. For instance, the index "1-2" refers to the correlation of the data collected by the first antenna and the second one. According to [27], a strong correlation between the field samples means that the scattered field information on the entire measurement domain can be faithfully and fully recovered from the measured data without adding more antennas (i.e., w/o increasing the NOAs). Otherwise, a weak correlation suggests to add more antennas to collect additional non-redundant information. Therefore, the plots in Fig. 6 confirm that, regardless the presence or not of the noise on the scattered data, the NCC when using 9 antennas is smaller than that with 20 antennas, thus setting up more receiving antennas can be beneficial to the inversion process.

*C. Effects of Adding Virtual Antennas*

While theoretically adding more antennas could be a good choice for enhancing the reconstruction accuracy, practical limitations (e.g., limited available space, high experimental costs, mutual coupling effects, and longer acquisition-times) often advise against the direct/physical increment of the NOAs. Moreover, it is worth pointing out that the measurement time is a critical key-performance indicator in real-time imaging applications. Starting from these considerations, the completion of the scattered field information with non-redundant samples has been obtained by exploiting the interpolation method that adds "virtual antennas" to the imaging system instead of physical devices. Of course, the non-redundant scattered field data should be sufficient enough for the virtual concept to be effective starting from the premise that a suitable NCC value is within the range 0.2-0.5. This choice ensures that, on the one hand, the scattered fields collected by the different antennas are distinct, on the other hand, they contain non-trivial information on the scattering domain [27].

Therefore, while the original noisy data have been stored in $\overline{E}_{9\times 9}^{sca}$ to generate, according to Sect. I.B, the updated data set $\overline{E}_{N\times N}^{sca}$, this latter has been processed by the inversion methods in Sect. I.C. Fig. 7 shows the behavior of the average SSIM value for different noise levels, while the RE plots are not reported since they remains relatively unaltered analogously to the case in Fig. 5.

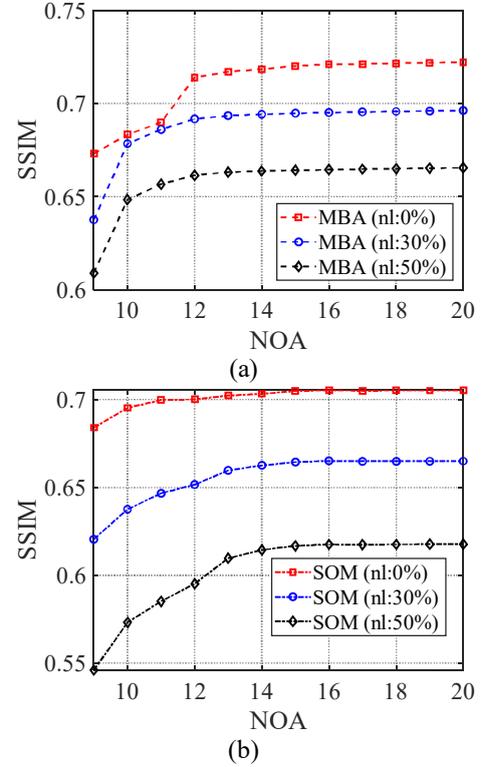

Fig. 7. Average SSIM of imaging varying with NOA (Interpolated) using different methods (a) MBA; (b) SOM.

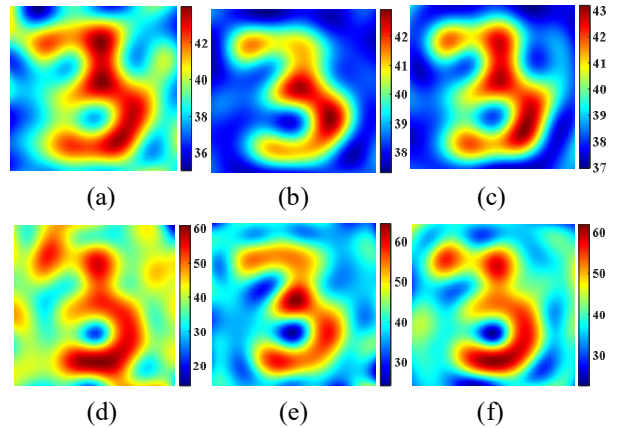

Fig. 8. Imaging comparison using MOM data for different NOA (a) 9×9 (Direct); (b) 20×20 (Interpolated); (c) 20×20 (Direct); (d) 9×9 (Direct); (f) 20×20 (Interpolated); (f) 20×20 (Direct) ((a), (b), and (c) are obtained using MBA method, while (d), (e), and (f) are obtained using SOM. The size of the imaging area is the area of DOI).

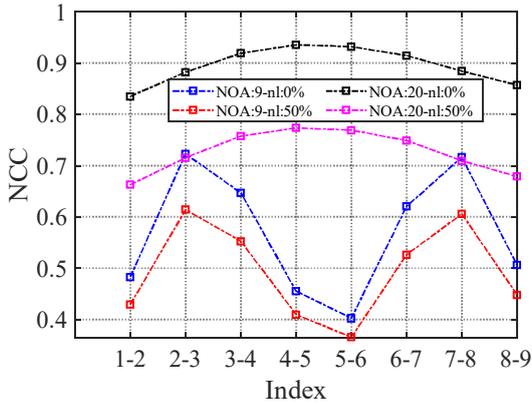

Fig. 6. Normalized correlation coefficient (NCC) under different noise levels.

It is interesting to note that, independently on the inversion method (i.e., the contrasts of the actual scatterers) and whatever the noise levels, the introduction of more virtual antennas leads to an improvement of the imaging accuracy. To point out also pictorially such an effect, the reconstructions of the digital "3" from noisy ($nl$ = 50%) data are reported in Fig. 8. One can observe that when NOA = 9 there are several artifacts in the DOI. By enlarging the scattered dataset either growing the NOAs or adding some "virtual antennas", the image of the DOI gets better and better.

## IV. VALIDATION AGAINST SEMI-EXPERIMENTAL DATA (HFSS SIMULATION)

While the numerical analysis in Sect. III gives us some insights on the possibility to improve the imaging performance of microwave imaging by increasing the NOAs, the use of ideal sources instead of real antennas is a limitation towards real implementations since it neglects the coupling effects and the system "noise" that cannot be faithfully modeled with only an additive Gaussian white noise. Therefore, the system in Fig. 1 has been emulated in the full-wave simulation software HFSS by assuming as transmitter/receiver a realistic Vivaldi antenna [42]. By considering a cylindrical background medium of height 0.5 [m] with relative permittivity $\varepsilon_b$ = 37.725 and loss tangent 0.148, the antenna sensors have been uniformly distributed along a circular path 0.12 [m] in radius. Since the HFSS full wave simulation of the 2.5-D imaging system is very time-consuming, the number of scatterers under test has been limited to 200 handwritten models, one half with relative permittivity randomly selected in the range [40, 50] (MBA-inversion), the other half within [50, 60] (SOM inversion). Moreover, the transmission coefficients among different antenna ports have been measured and then transformed into scattered field data by using the calibration method in [42].

The inversion results are summarized in Tab. I in terms of the average values of the SSIM and the RE for different arrangements of the NOAs. Please note that the notation $9 \rightarrow 20$ indicates that NOA = 9 actual antennas are expanded into N = 20 antennas by adding 11 virtual measurement locations. As it can be inferred, the inversion accuracy when processing data collected by NOA = 9 actual sensors is worse than that with virtual antennas (e.g., SSIM: 0.554 [NOA = 9] vs. 0.626 [N = 20] – MBA-inversions; SSIM: 0.672 [NOA = 9] vs. 0.697 [N = 20] – SOM-inversions). Moreover, the use of a larger set of actual antennas (i.e., NOA = 20) does not give the same improvements (e.g., SSIM: 0.626 [N = 20] vs. 0.493 [NOA = 20] – MBA-inversions; SSIM: 0.697 [N = 20] vs. 0.692 [NOA = 20] – SOM-inversions) even the opposite in the case of moderate scatterers (e.g., SSIM: 0.493 [NOA = 20] vs. 0.554 [NOA = 9], RE: 7.701 [NOA = 20] vs. 0.041 [NOA = 9] – MBA-inversions). This latter outcome is caused by the fact that the weaker field from moderate scatterers is submerged by the "noise" of the strong coupling among the antenna ports, which increases by adding more real antenna elements.

The quantitative indications drawn from Tab. I are confirmed by the reconstructions in Fig. 9.

## V. VALIDATION AGAINST EXPERIMENTAL RESULTS

TABLE I.
AVERAGE RESULTS OF SSIM AND RE USING HFSS DATA FOR DIFFERENT NOAS

| Method | MBA | | | SOM | | |
|---|---|---|---|---|---|---|
| NOA | 9 | 9→20 | 20 | 9 | 9→20 | 20 |
| SSIM | 0.554 | 0.626 | 0.493 | 0.672 | 0.697 | 0.692 |
| RE | 0.041 | 0.036 | 7.701 | 0.057 | 0.056 | 0.066 |

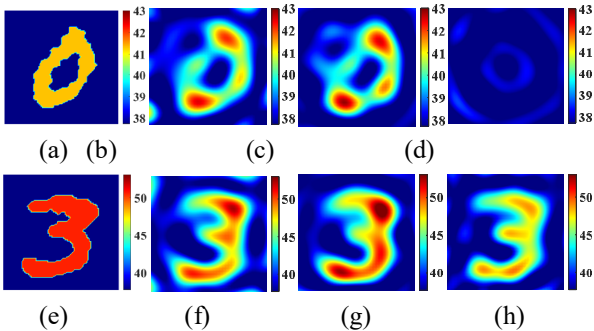

Fig. 9. Imaging comparison using HFSS data for different NOA (a) Ground truth; (b) 9×9 (Direct); (c) 20×20 (Interpolated); (d) 20×20 (Direct); (e) Ground truth; (f) 9×9 (Direct); (g) 20×20 (Interpolated); (h) 20×20 (Direct) ((b), (c), and (d) are obtained using MBA method, while (f), (g), and (h) are obtained using SOM. The size of the imaging area is the area of DOI).

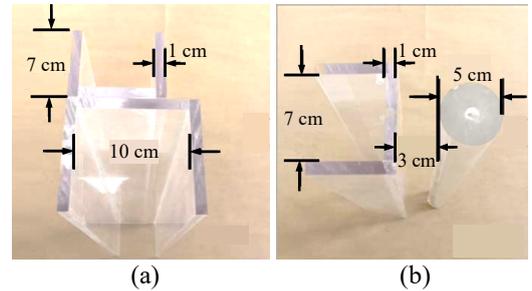

Fig. 10. Two scatterers with different shapes machined from organic glass (a) C-C model; (b) C-O model.

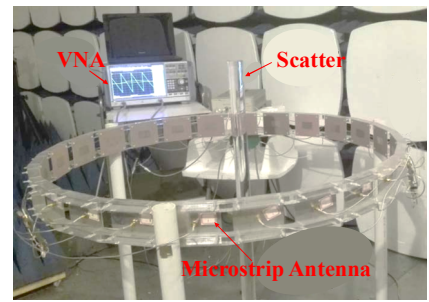

Fig. 11. A practical microwave imaging system.

For the experimental validation, the two scatterers in Fig. 10 have been fabricated from organic glass [43]. The first object features two C-shaped structures and it will be referred hereafter as the "C-C model". The second one is a combination of a C-shaped and a circular structure ("C-O model"). The height of both scatterers is 100 [cm], while the relative permittivity has been estimated to be $\varepsilon_r \approx 3$. The data acquisition system is shown in Fig. 11 and it consists of a VNA and a circular microstrip antenna array working at 2.4 [GHz]. There are 24 antennas evenly distributed along a circle 0.565 [m] in radius and each antenna serves as both transmitter and receiver. The angular spacing between two transmitting antennas has been set to 30°. This means that 12 of the 24 sensors are used as transmitters. Moreover, at each transmission, the remaining 21 antennas (the two adjacent antennas have been excluded owing to the pronounced coupling effects) operate as receivers. Consequently, the size of the actual measured scattered field matrix turns out to be $21 \times 12$ -sized, $\bar{E}^S_{21 \times 12}$.

Starting from the computation of the DOFs of the two scattering examples (DOFs = 11 - "C-C model"; DOFs = 9 - "CO model"), firstly the interpolation method has been used to convert the measured matrix $\bar{E}^S_{21 \times 12}$ into the two square matrices $\bar{E}^S_{11 \times 11}$ and $\bar{E}^S_{9 \times 9}$, which have been extended to the matrix $\bar{E}^S_{20 \times 20}$ with the FDZP method in Sect. I.B.

Fig. 12 show the reconstruction when processing the different datasets. As expected, the reconstructions from the NOAs equal to the number of DOFs (i.e., $\bar{E}^S_{11 \times 11}$ - "C-C model"; $\bar{E}^S_{9 \times 9}$ - "CO model") present several artifacts, while the use of more field samples (i.e., $\bar{E}^S_{20 \times 20}$) significantly improve the imaging performance by making the contours of the scatterers clearer and filtering out some wrong details, as well. It is also worth noticing that the direct processing of the data collected by the experimental system (i.e., $\bar{E}^S_{21 \times 12}$) reaches good results. This is not in contradiction with the outcomes in Sect. IV since here the port coupling effects have been minimized by avoiding the measurements from the receiving antennas adjacent to the transmit one. For completeness, Tables II-III give the average values of the error indexes.

TABLE II.
AVERAGE RESULTS OF SSIM AND RE USING MBA METHOD

| | C-C model | | | C-O model | | |
|---|---|---|---|---|---|---|
| NOA | $\bar{E}^S_{11 \times 11}$ | $\bar{E}^S_{20 \times 20}$ | $\bar{E}^S_{21 \times 12}$ | $\bar{E}^S_{9 \times 9}$ | $\bar{E}^S_{20 \times 20}$ | $\bar{E}^S_{21 \times 12}$ |
| SSIM | 0.600 | 0.685 | 0.712 | 0.526 | 0.677 | 0.724 |
| RE | 0.317 | 0.293 | 0.287 | 0.222 | 0.210 | 0.220 |

TABLE III.
AVERAGE RESULTS OF SSIM AND RE USING SOM

| | C-C model | | | C-O model | | |
|---|---|---|---|---|---|---|
| NOA | $\bar{E}^S_{11 \times 11}$ | $\bar{E}^S_{20 \times 20}$ | $\bar{E}^S_{21 \times 12}$ | $\bar{E}^S_{9 \times 9}$ | $\bar{E}^S_{20 \times 20}$ | $\bar{E}^S_{21 \times 12}$ |
| SSIM | 0.661 | 0.718 | 0.716 | 0.537 | 0.716 | 0.782 |
| RE | 0.291 | 0.282 | 0.286 | 0.247 | 0.220 | 0.215 |

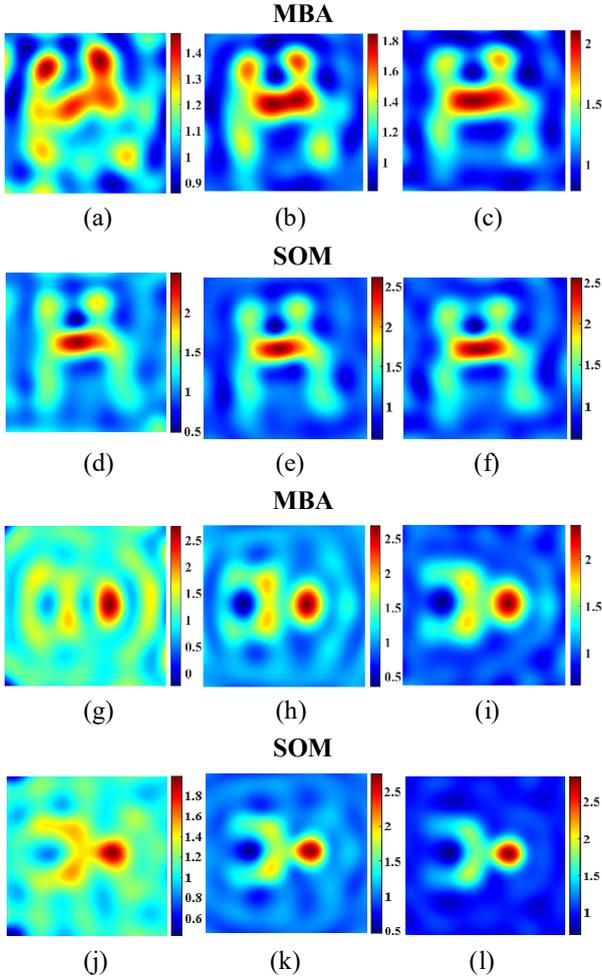

Fig. 12. Imaging comparison using scattered field matrices with different dimensions (a) $\bar{E}^S_{11 \times 11}$; (b) $\bar{E}^S_{20 \times 20}$; (c) $\bar{E}^S_{21 \times 12}$; (d) $\bar{E}^S_{9 \times 9}$; (e) $\bar{E}^S_{20 \times 20}$; (f) $\bar{E}^S_{21 \times 12}$ ((a), (d), (g)and (j) are calculated by using $\bar{E}^S_{11 \times 11}$ or $\bar{E}^S_{9 \times 9}$, (b), (e), (h)and (k) are calculated by using $\bar{E}^S_{20 \times 20}$, (e), (f), (i) and (l) are calculated by using $\bar{E}^S_{21 \times 12}$, and the size of the imaging are is 0.25 [m]×0.25 [m]).

## VI. CONCLUSION

This paper has investigated the dependence of the inversion accuracy on the NOAs in real microwave imaging systems featuring non-ideal behaviors. It has been proved that increasing the NOAs can be beneficial since the correlation of the scattered data collected by neighboring receivers decreases with the "environmental/measurement noise". Because of the limitations of real imaging system to arrange more antennas in the acquisition setup, an interpolation method based on the FDZP has been proposed to yield non-redundant scattered field samples by adding virtual antennas. The effectiveness and the reliability of the proposed approach has been assessed by processing, with reliable state-of-the-art inversion techniques synthetic, semi-experimental, and experimental datasets from 2D and 2.5D scattering scenarios. Future works will be aimed at extending such an approach to other applications (e.g., buried object detection and NDE/NDT) as well as to 3D scenarios.

## VII. ACKNOWLEDGMENT

The authors would like to thank very much Prof. L. Ran (College of Information Science and Electronic Engineering, Zhejiang University China) and Prof. K. Xu (School of Electronics and Information Engineering, Hangzhou Dianzi University, China) for their great help in acquiring and processing the measurement data.